\documentclass[aps,pra,12pt,nofootinbib,superscriptaddress,longbibliography,showpacs]{revtex4-1}

\usepackage[utf8]{inputenc}
\usepackage{amssymb,amsmath,amsfonts,amsthm}
\usepackage{verbatim}
\usepackage{enumerate}
\usepackage{tensor}
\usepackage{fancyhdr}
\usepackage{graphicx}
\usepackage{hyperref}

\usepackage{color}
\definecolor{dred}{rgb}{.8,0.2,.2}
\definecolor{ddred}{rgb}{.8,0.5,.5}
\definecolor{dblue}{rgb}{.2,0.2,.8}

\theoremstyle{plain}
\newtheorem{theorem}{Theorem}   

\newtheorem{corollary}[theorem]{Corollary}

\theoremstyle{definition}
\newtheorem{definition}[theorem]{Definition}
\newtheorem{example}[theorem]{Example}
\newtheorem{remark}[theorem]{Remark}

\newcommand{\bra}[1]{\mbox{$\langle #1|$}}
\newcommand{\ket}[1]{\mbox{$|#1\rangle$}}

\newcommand{\ketbra}[2]{\mbox{$|#1\rangle\langle #2|$}}

\newcommand{\gate}[1]{\ensuremath{\text{\sc #1}}}
\newcommand{\COPY}[1][]{\ensuremath{\gate{COPY}_{#1}}}

\DeclareMathOperator{\Tr}{Tr}

\DeclareMathOperator{\Aut}{Aut} 
\DeclareMathOperator{\End}{End} 

\newcommand{\I}{\openone}     
\newcommand{\R}{{\mathbb R}}  
\newcommand{\C}{{\mathbb C}}  
\newcommand{\hilb}[1]{\ensuremath{\mathcal{#1}}} 
\newcommand{\swap}{{\sf{SWAP}}}
\newcommand{\ie}{i.e.}

\def\1#1{{\bf #1}}
\def\2#1{{\cal #1}}
\def\3#1{{\sl #1}}
\def\4#1{{\tt #1}}
\def\5#1{{\sf #1}}
\def\6#1{{\mathfrak #1}}
\def\7#1{{\mathbb #1}}

\newcommand{\be}{\begin{equation}}
\newcommand{\ee}{\end{equation}}

\newcommand{\Figref}[1]{Figure \ref{#1}}

\begin{document}

\newlength{\diagwidth}
\setlength{\diagwidth}{10cm}



\title{Tensor Network Methods for Invariant Theory}

\author{Jacob Biamonte}
\email{jacob.biamonte@qubit.org}
\affiliation{Institute for Scientific Interchange,
Via Alassio 11/c, 10126
Torino, Italy}
\affiliation{Centre for Quantum Technologies, National University of Singapore
Block S15, 3 Science Drive 2,
Singapore 117543
}

\author{Ville Bergholm}
\email{ville.bergholm@iki.fi}
\affiliation{Institute for Scientific Interchange,
Via Alassio 11/c, 10126
Torino, Italy}
\affiliation{Department of Chemistry, Technische Universität München, D-85747
Garching, Germany}

\author{Marco Lanzagorta}
\email{marco.lanzagorta@nrl.navy.mil} 
\affiliation{US Naval Research Laboratory,
4555 Overlook Ave. SW,
Washington DC
 20375
}


\begin{abstract}
Invariant theory is concerned with functions that do not change under
the action of a given group.  Here we communicate an approach based on tensor
networks to represent polynomial local unitary
invariants of quantum states.  This graphical approach provides an alternative
to the polynomial equations that describe invariants, which often contain a
large number of terms with coefficients raised to high powers.
This approach also enables one to use known methods from tensor
network theory (such as the matrix product state factorization) when studying
polynomial invariants. As our main example, we consider invariants of matrix
product states. We generate a family of tensor contractions resulting in a
complete set of local unitary invariants that can be used to express the Rényi
entropies.  We find that the graphical approach to representing invariants can
provide structural insight into the invariants being contracted,
as well as an alternative, and sometimes much simpler, means to study polynomial
invariants of quantum states. In addition, many tensor network methods, such as
matrix product states, contain excellent tools that can be applied  
in the study of invariants.  
\end{abstract}

\maketitle

In quantum physics, polynomial invariants typically arise in the study of
various entanglement-related properties of quantum states~\cite{grassl1998, makhlin2002, entinv3, qubits3, entinv1,2011PhRvA..83f2308W}
and gates~\cite{PhysRevA.63.062309,qic2006}.
In this paper we present a variant of the graphical tensor calculus of
Penrose~\cite{Penrose} for the purpose of representing and computing polynomial
invariants of arbitrary quantum states.

The graphical approach of tensor network diagrams\footnote{We assume readers are
familiar with the basics of tensor networks, although we
will review them in Section \ref{sec:penrose}.  Readers seeking the basics of
tensor networks could consult \cite{CTNS, BB11, DBJC11} and those interested in
tensor network algorithms and applications to physics could consult
\cite{2011AnPhy.326...96S, MPSreview08,TNSreview09}.} provides an
alternative to the polynomial equations that describe invariants, which often
contain a large number of terms with coefficients raised to high powers. It also
enables one to use established methods from tensor network theory (such as
matrix product state factorizations) to study polynomial invariants.  We find in
our examples that the underlying mathematical structure of the physics described
by the invariants is reflected in the structure of the resulting tensor
networks.  By using specific graphical rewrite rules, our methods enable one 
to contract and simplify the tensor network representing any given
polynomial invariant of a bipartite pure state
to the point where the network is succinctly
expressed in terms of Schmidt coefficients.  This serves as a graphical
proof of the invariance of the quantity represented by the network, as
well as a conceptual aid geared towards understanding
the meaning behind the invariants. The graphical method introduced to factor
matrix product states is slightly different than known approaches in the
literature---see for example, the matrix product states review
\cite{2011AnPhy.326...96S} and \cite{MPSreview08,TNSreview09}. These differences
are important for
our purposes.  

The area of tensor network states and tensor network algorithms is a rapidly
growing area of physics which studies (in part) the most efficient way to
represent quantum states and to discover key properties of quantum systems. 
One of the main methods inside this framework is the matrix product state (MPS)
representation~\cite{2003PhRvL..91n7902V, MPSreview08,TNSreview09, 2011AnPhy.326...96S, 2006quant.ph..8197P}. 
This method is not well
known outside of physics.  We find it to be well suited to study invariants
and think that others working in the area of invariant theory (even outside of
physics) will find the matrix product state factorization useful.  We therefore
hope that the present paper can help bridge this gap between these communities
and foster further cross-pollination between invariant theory and tensor
network states.
For related work published after the preprint version of our study, see~\cite{CM12, turner2013}.

At the heart of MPS is the tensor network description of repeated bipartitions
of a quantum state.  By capturing the singular value decomposition in a
tensor network where all internal components have clearly defined
algebraic properties, we present some small improvements in the graphical
tensor calculus used to describe matrix
product states, as well as invariants in general.
In this regard, our results on matrix product states take an
important first step in uniting invariant theory with tensor network
states.  The key example we consider here is showing the utility of tensor
network methods for matrix product factorizations of quantum states into
bipartitions.  

Apart from the intuition found in representing states as MPSs, one might
also employ tensor network
algorithms~\cite{MPSreview08,TNSreview09,2006quant.ph..8197P,tommy,
2012PhRvB..86s5114S} to
design and contract invariants of interest to physics. Tensor network methods
also offer a valuable conceptual aid to understanding how the
numerical value of an invariant relates to properties
of the state.

We begin by recalling the fundamental notions of the tensor calculus in Sections
\ref{sec:penrose}, \ref{sec:wireduals} and~\ref{sec:diagramaticSVD}.
This leads to the diagrammatic SVD, which is
used in Section~\ref{sec:dmps} to factor a given quantum state into a
matrix product state.  We then connect the tensor calculus to
polynomial invariants in Section~\ref{sec:qubit-invariants}.  Before
concluding, we also consider the application of invariants to calculate
entropies and entanglement measures.

\section[Penrose Graphical Notation]{Penrose Graphical Notation for Tensor Networks} \label{sec:penrose}

Penrose graphical notation~\cite{Penrose} is a diagrammatic notation
for tensor networks.  This notation is becoming well known inside the tensor
network algorithms community (see for example \cite{2007PhRvA..76e2315G} for an
early use of the graphical notation to describe matrix product states). 
It can make the manipulation of complicated tensor networks much
easier and more intuitive.  Contributions on the topic we found influential can 
be found in \cite{Laf92, Laf95, boolean03, 2009arXiv0903.0340B, prehistory}. In
our previous work, we have adapted the graphical notation and surrounding
methods to describe generalized quantum circuits~\cite{BB11}, tensor network
states \cite{CTNS, BB11, DBJC11}, open quantum systems \cite{WBC11, MB12} as
well as decidability in algorithms based on tensor contractions \cite{MoB12}.  

In the string diagram notation, a tensor is a graphical shape with a number of input
legs (or ``arms'') pointing up, and output legs pointing down.\footnote{Often, to conserve space, the diagrams are rotated $90$~degrees
counterclockwise. In practice this should be obvious from the context.}
Individual arms as well as individual legs each independently correspond to an index. For example,
\be
 \includegraphics[width=0.6\diagwidth]{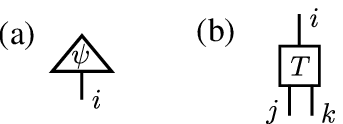}
\ee
diagram (a) represents the tensor~$\psi_i$ and (b) the
tensor~$T^i_{~jk}$.  A tensor with $n$ indices up and
$m$ down is called a valence-$(n,m)$ tensor and sometimes a
valence-$k$ tensor for $k=n+m$.

In quantum physics parlance one introduces a computational basis
and expands the tensors in it; in which case $T^i_{~jk}$ is understood
not as abstract index notation but as the actual components of the tensor:
\be 
T = \sum_{ijk} T^i_{~jk}\ket{jk}\bra{i}.
\ee 
In practice there is little room for confusion however.

There are three special ``wire tensors'' that
play the role of the metric tensor.\footnote{We will always work in a flat Euclidean space,
which renders the metric tensors trivial.}
They are given diagrammatically as
\be
\includegraphics[width=1.2\diagwidth]{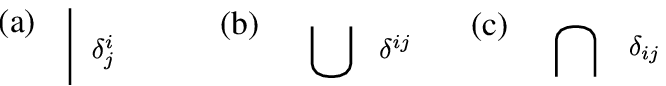}
\ee
The identity tensor (a) is used for index contraction
by connecting the corresponding legs, and the cup (b) and cap (c) are 
metric tensors used 
for raising and lowering indices.
Expanding them in the computational basis we obtain
\begin{align}
\label{eqn:delta}
\I &= \sum_{ij}\delta\indices{^i_j}\ket{j}\bra{i} =\sum_k\ket{k}\bra{k},\\
\label{eqn:delta-effect}
\bra{\cup} &= \sum_{ij} \delta\indices{^{ij}}\bra{ij} = \sum_k\bra{kk}, \quad \text{and}\\
\label{eqn:delta-bell}
\ket{\cap} &= \sum_{ij} \delta\indices{_{ij}}\ket{ij} = \sum_k\ket{kk}.
\end{align}

\begin{figure}
\includegraphics[width=1.0\diagwidth]{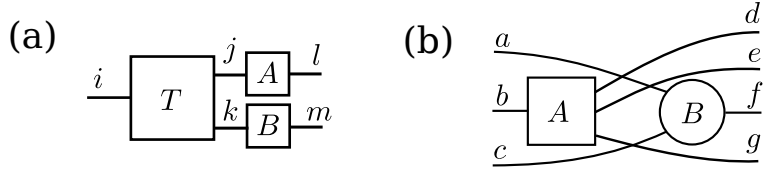}
\caption{Illustration of the graphical notation.
(a) Contraction of tensor $T$ with tensors $A$ and $B$ amounts to joining indices:
$T^i_{~jk}A^j_{~l}B^k_{~m}$.
(b) Permutation of indices by crossing wires: $A^{b}_{~deg}B^{ac}_{~~f}$.
\label{fig:contractions}
} 
\end{figure} 

\begin{enumerate}
\item[(i)]
One can raise and subsequently lower an index or vice versa, which
amounts essentially to doing nothing at all.  This scenario
is captured diagrammatically by the so called \emph{snake} or
\emph{zig-zag equation}
\be
 \includegraphics[width=0.6\diagwidth]{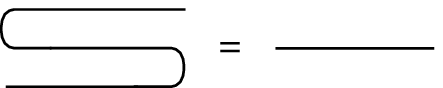}
\ee
together with its mirror image~\cite{Penrose}.
In tensor index notation, it is expressed succinctly as $\delta^{ij} \delta_{jk} = \delta^i_{~k}$.  

\item[(ii)]
Crossing two wires (as in diagram (a) below) is equivalent to swapping the relative order of the
corresponding vector spaces.
\be
 \includegraphics[width=0.8\diagwidth]{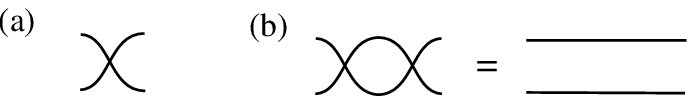}
\ee
(b) illustrates that the swap operation is self inverse.
It may be written as $\swap^{ij}_{~~kl}=\delta^i_{~l}\delta^j_{~k}$.

\item[(iii)]
The trace in the graphical calculus is given by appropriately joining wires to close loops.
\end{enumerate}

Together the cups and caps give rise to a correspondence between different types of maps and states.
We call the duality induced by bending and exchanging wires \emph{Penrose duality}.

\section[Penrose Wire Bending Duality]{Penrose Wire Bending Duality}\label{sec:wireduals}
Now we will consider
the set of operations formed from bending tensor wires forwards or
backwards using cups and caps, as well as exchanging wires using
\swap{}.  We can conceptualize this set of transforms acting on a
tensor as amounting essentially to matrix reshapes. From the snake
equation, action with a cup or a cap
is invertible and \swap{} is self inverse.  This implies that all
possible configurations of a tensor's wires obtained using these
operations can be considered equivalent.
We will start with an example.

\begin{example}\label{ex:index} 
  Given a tensor $T^{i}_{~j}$ with fixed labels $i,j$ one uses cups and caps to rearrange index elevations, arriving at 
\be 
T^{i}_{~j}, ~ T^{ij}, ~ T_{ij}, ~ T_{i}^{~j}
\ee 
Using the \swap{} operation one reorders the horizontal position of $i$ and $j$.  Then applying cups and caps yields 
\be 
T^{j}_{~i}, ~  T^{ji}, ~ T_{ji}, ~ T_{j}^{~i}
\ee 
for a total of eight possible reshapes. 
\end{example}

For an $n$ index tensor, each index can be either up or down, yielding
$2^n$ possibilities.  The symmetry group formed by \swap{} is of order
$n!$ and acts to arrange the horizontal position of the $n$ legs of a
tensor, yielding
(provided we distinguish forms of the type $T_{i}^{~j}$ and $T^{j}_{~i}$ in Example~\ref{ex:index})
$n! \cdot 2^n$ different ways to reorder the $n$ indices of a tensor.

\begin{remark}[Ordering operators by numbers of inputs and outputs]\label{remark:ordering-index}
In the previous example, we considered $T_{i}^{~j}$ (b) and $T^{j}_{~i}$ (a) as distinct. This is illustrated in (a) and (b) as follows.  
 \be
 \includegraphics[width=0.7\diagwidth]{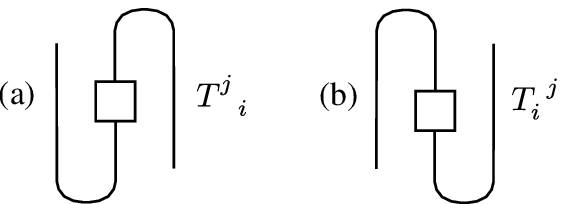}
\ee
This provides an example of an awkward property of the inherently
one-dimensional Dirac notation.  Both (a) and (b) represent the same map,
but when we write them in a basis, consistency dictates that one will
expand in the basis $\bra{i}\otimes\ket{j}$.

With the equivalence explained in Remark \ref{remark:ordering-index} in mind, we
note that the tensor $T^{i}_{~j}$ from
Example \ref{ex:index} actually has six unique reshapes, as two pairs
of reshapes are diagrammatically equivalent.  In other words, in (b) below we have that $T^i_j = T^i_{~j} = T^{~i}_j$ and for (e) we have the equality $T^j_i = T^j_{~i} = T^{~j}_i$. 
 \be
 \includegraphics[width=1.0\diagwidth]{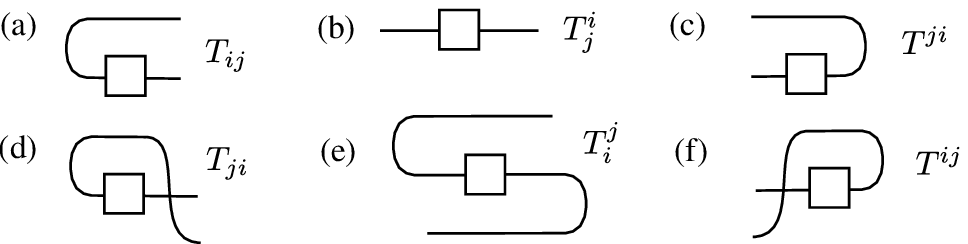}
\ee
\end{remark}
Together, we call these reshapes the \emph{natural tensor symmetry class}.
More generally one finds the number of diagrammatically unique
reshapes of a tensor by
(i) counting the number of possible ways it can have its wires bent, either forward or backwards using the
cups and caps, and (ii) the number of ways a tensor
can have its arms and/or legs reordered. We arrive at the following
result:
\begin{theorem}[Natural tensor symmetry class]
\label{theorem:symmetry-class}
The arms and legs of a tensor $\Gamma^{ij\cdots k}_{qr\cdots s}$ with
$n$ input and output legs in total can be rearranged in  
$(n+1)!$ different ways.
\end{theorem}

\section{Diagrammatic SVD}\label{sec:diagramaticSVD}

In this section, we introduce the diagrammatic representation of
the singular value decomposition (SVD).
Later it will be used to simplify
invariants obtained through network contraction,
and iterated to obtain a matrix product state (MPS) description for a
pure state.

The SVD factors tensors into well defined building blocks
with simplistic interaction properties:
(i) a valence-one tensor storing singular values,
(ii) a valence-three-\COPY~tensor used to create a diagonal map, and
(iii) a pair of valence-two unitary gates.
\COPY-tensors have been studied in the setting of the Penrose
tensor calculus, in work dating back at least to Lafont~\cite{Laf92, Laf95} --- see also \cite{boolean03, BB11, CTNS}.

\begin{definition}[\COPY{} tensor]
\label{def:copy-prop}
The $m$-to-$n$ \COPY{} tensor is defined in the computational basis as
\be
\COPY[m \to n] := \sum_{k=0}^{d-1} \ketbra{\underbrace{k \cdots k}_{n}}{\underbrace{k \cdots k}_{m}}.
\ee
It is named accordingly because
connecting a basis state $\ket{k}$ to
any of its input or output wires collapses the sum and breaks
the tensor up into unconnected copies of~$\ket{k}$ and~$\bra{k}$.
As with classical circuits, in the diagrammatic tensor notation \COPY[m \to n] is represented by a
simple black dot~$\bullet$ with $m$~input and $n$~output legs.
Since all the legs of a \COPY{} tensor are identical, and the
inputs can be converted to outputs and vice versa simply by using cups and
caps, keeping track of the direction of the legs is not important as
long as they are connected to other tensors.
This is reflected in the notation.
For a brief enumeration of the algebraic properties of the \COPY{} tensor, see~\cite{BB11}.
\end{definition}


\begin{theorem}[Diagrammatic SVD]
\label{theorem:diagrammatic-SVD}
Any valence-two tensor $f: A \to B$ can be
factored into a non-negative, unique valence-one tensor $\Sigma$,
a valence-three \COPY{} tensor,
unitary valence-two tensors $U$~and~$V$,
and a diagonal valence-two dimension changer tensor $Q$ 
when necessary:
\be
\includegraphics[width=0.9\textwidth]{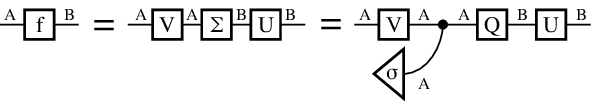}\label{fig:SVD-s}
\ee
The dimension changing tensor $Q: A \to B$ has 1's on the diagonal and
zero entries otherwise.
\begin{proof}
The SVD of $f$ is
\be
f = U \Sigma V,
\ee
where $U: B \to B$ and $V: A \to A$ are unitary operators and
$\Sigma: A \to B$ is diagonal in the computational basis, with the
(necessarily non-negative) singular values $\sigma_i$ of $f$
along the diagonal.
$\Sigma$ can be written as
\be 
\Sigma
= \sum_{j=0}^{d-1} \sigma_j \ket{j}_B \bra{j}_A
= \underbrace{\sum_{i=0}^{d-1} \ket{i}_B \bra{i}_A}_{Q_{AB}}
\underbrace{\sum_{j} \ket{j}_A\bra{jj}_A}_{\COPY[2 \to 1]}
\underbrace{\sum_k \sigma_k \ket{k}_A}_{\sigma}
\qquad (\sigma_k \ge 0),
\ee
where $d = \min(\dim A, \dim B)$.
We have expressed $\Sigma$ as a contraction of an
valence-one tensor~$\sigma$ with a \COPY{} tensor.
The non-square tensor $Q_{AB}$ is only necessary if $A$ and $B$
have different dimensions.
\end{proof}
\end{theorem}

\begin{corollary}[Diagrammatic Schmidt decomposition]
\label{ex:diagrammatic-Schmidt}
Given a bipartite state $\ket{\psi} \in A \otimes B$, we use the snake equation to
convert it into a linear map~$f:A \to B$ (inside the dashed region
below):\footnote{
This has also been understood as a diagrammatic form of map-state duality underlying bipartite entanglement evolution \cite{MB12}.}
\be
\label{fig:SVD-Schmidt2}
\includegraphics[width=0.9\textwidth]{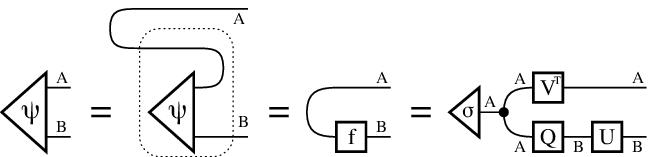}
\ee
Now we apply the SVD as in
Theorem~\ref{theorem:diagrammatic-SVD}.
Diagram reorganization leads to the diagrammatic Schmidt decomposition
of~$\ket{\psi}$.
The singular values in $\sigma$ now correspond to the Schmidt coefficients.
\end{corollary}

The network topology of the diagrammatic Schmidt decomposition can be
used to study the entanglement properties of the bipartite state~$\ket{\psi}$.

\begin{example}[Entanglement topology]
\label{example:entanglement-topology}
The topology of a bipartite state
$\ket{\psi} = \sum_i \sigma_i \ket{\varphi_i}\ket{\phi_i}$
depends on the
singular values in the triangular tensor
$\ket{\sigma} = \sigma_0 \ket{0} +\sigma_1 \ket{1} +\ldots
+\sigma_{d-1} \ket{d-1}$.
\be
\includegraphics[width=0.8\diagwidth]{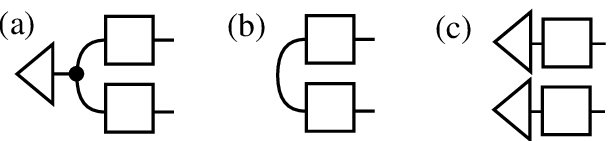}
\ee
The general diagram state (a) takes the form (b) iff the Schmidt coefficients
all have the same value.
Now $\ket{\sigma}$ is proportional to a unit for the \COPY-tensor, and
the tensor structure is converted to a smooth wire, yielding
the maximally entangled case.
The most significant topology change (c) occurs when the input state to
the \COPY{} tensor is a single basis state
$\ket{\sigma}=\ket{0}$. 
As this is a copy-point for the
\COPY-tensor, it breaks into two copies of~$\ket{\sigma}$ and
separates the diagram into two halves, illustrating the fact that the state is factorizable.
\end{example}

\section{Diagrammatic Matrix Product States}\label{sec:dmps}

We will now consider matrix
product states~(MPS), an iterative method to factor quantum states
into a linear chain of tensors (see~\cite{MPSreview08,TNSreview09, 2011JSP...145..891E} for
a recent review and~\cite{BB11} for work considering the category
theory behind MPS).  The reason this factorization is called a 1D method is because it is known to 
describe a class of 1D systems efficiently, and because the
factorization results in a 1D chain (for a discussion of other
factorizations and the connection to geometry see for instance 
\cite{2011JSP...145..891E}).
Without loss of generality, we will apply the MPS method to a
four-party state, and explain the procedure in terms of three distinct
steps.  

\begin{remark}[Method summary]
MPS correspond to an iterative factorization method for quantum states. 
The key idea is a recursive application of the singular value decomposition
(SVD). It begins by first selecting a bipartition and then applying the SVD. 
If either of the initial bipartitions can themselves be bipartitioned, the SVD
is applied again.  This results in a 1D tensor network representation of the
state, as described in further detail below.
\end{remark}

Consider a quantum state, expressed as a triangle in the
Penrose graphical notation with a label $\textbf{1}$ inside and open
legs labeled $i,j,k,m$.
\be
 \includegraphics[width=0.2\diagwidth]{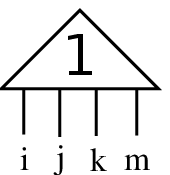}
\ee

\begin{figure}[t]
\includegraphics[width=1.6\diagwidth]{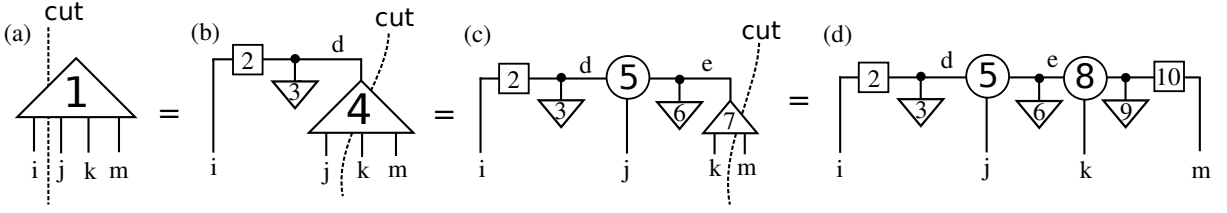}
\caption{MPS factorization steps (diagrammatic description of steps I,
  II and III). The quantum state (a) is iteratively factored into the
  1D matrix product state (d). This procedure readily extends to
  $n$-body states.
\label{fig:MPS-decomposition}
}
\end{figure}

(Step I). We will create a bipartition comprising a first collection,
containing only leg $i$ and a second, containing legs $j,k,m$.  We
will then apply the diagrammatic SVD across this partition.
The partition is illustrated with the dashed \emph{cut} below in (a).
Diagram (b) results from applying the diagrammatic SVD across this
partition, factoring the original state labeled $\textbf{1}$
into a valence-two unitary box labeled $\textbf{2}$, a valence-one
triangle containing the singular values labeled $\textbf{3}$, and a
valence-four triangle labeled $\textbf{4}$, all contracted with a
\COPY-tensor, as illustrated.  A new internal label (d) for the wire
connecting the \COPY-tensor to the valence-four triangle
($\textbf{4}$) was introduced for clarity. See also Fig.~\ref{fig:MPS-decomposition}(a,b).
\be
 \includegraphics[width=0.6\diagwidth]{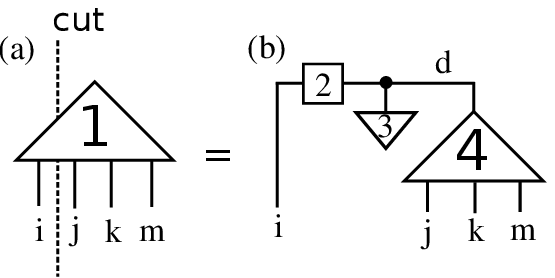}
\ee

\begin{remark}[Isometric internal tensors]
\label{remark:isometry}
The valence-four triangle tensor labeled $\textbf{4}$ above arises from contracting
a unitary map with a dimension changing tensor $Q$ (see (a) below).
The input leg shown is labeled d.  The other legs are contracted with
a fixed basis state $\ket{0}$, from the SVD in (a) above.
From the unitarity property, the isometry property follows, as illustrated
graphically in (c) and (d) below.
 \be
  \includegraphics[width=1.4\diagwidth]{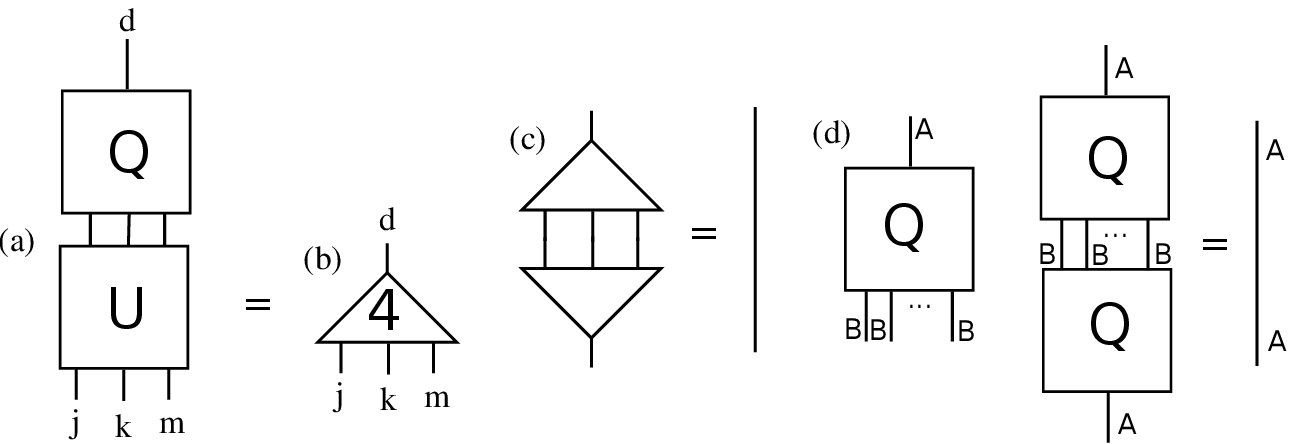}
 \ee
\end{remark}

(Step II). To illustrate the next step in the factorization, we will
remove the tensor labeled~$\textbf{4}$ by breaking the wire connecting
it to the \COPY-tensor (a).  We will then partition this separate
tensor into two halves, one containing wires $d, j$ the other half
wires $k,m$.  This partition is illustrated by placing a dashed line
(labeled cut) in (a).   We arrive at the structure in (b), which
we have explained in the first step. See also Fig.~\ref{fig:MPS-decomposition}(b,c).
\be
 \includegraphics[width=0.6\diagwidth]{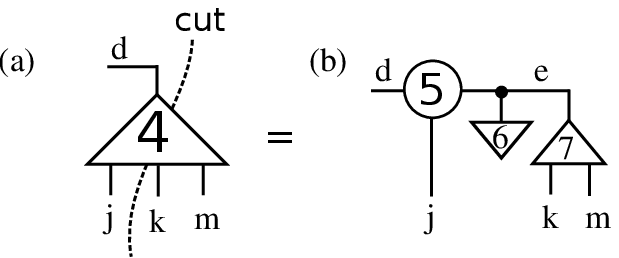}
\ee

\begin{remark}[An elementary property of tensor network manipulation]\label{remark:elementary-prop-1}
It is a fundamental property of tensor network theory that one can
remove a portion of a network, alter this removed portion of the
network without changing its function, and replace it back into the
original network, leaving the function of the original network
intact.
\end{remark}

(Step III).  In the third and final step of the MPS factorization
applied to this four-party example, following remark
\ref{remark:elementary-prop-1} we first place the tensor we have
factored in the second step, back into the original network from the
first step, see (a) below.  We then repeat the second step, applied to
the triangular isometry tensor, labeled internally with a
$\textbf{7}$.  This results in the factorization appearing in (b).
See also Fig.~\ref{fig:MPS-decomposition}(c,d).
\be
 \includegraphics[width=1.0\diagwidth]{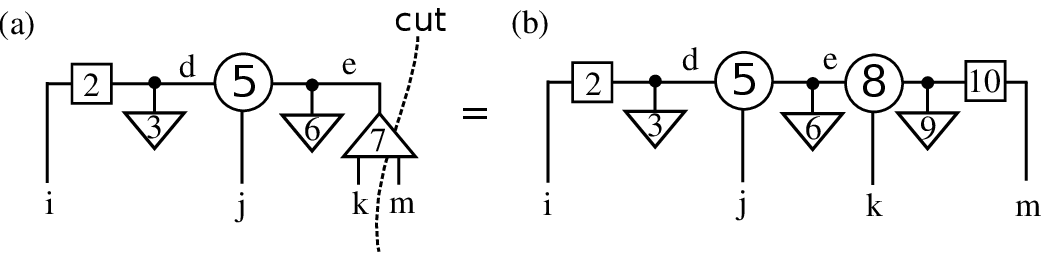}
\ee
The iterative method continues in the same fashion as the first three
steps, resulting in a factorization of an $n$-party state.  A summary
of the MPS factorization applied to a four-party state is shown in
Figure~\ref{fig:MPS-decomposition}.

\begin{figure}
\includegraphics[width=1.6\diagwidth]{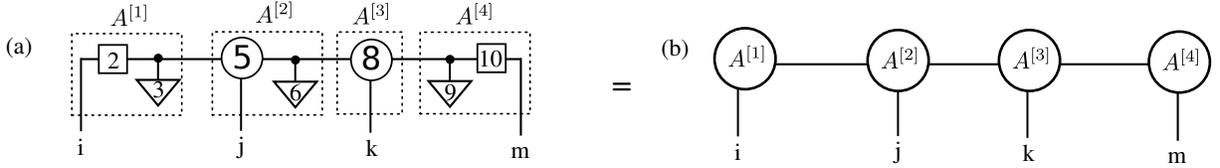}
\caption{Conversion from our notation (a), to conventional MPS
  notation (b). The factorization methods we have presented here and
  elsewhere \cite{CTNS, BB11} allow one to ``zoom in'' and expose
  internal degree of freedom (a) or ``zoom out'' and expose high-level
  structure (b). The equational representation of the MPS in (b) is
  given in \eqref{eqn:MPS}.
\label{fig:MPS-summary}
}
\end{figure}

(Summary). We will now consider Figure \ref{fig:MPS-summary}, which
summarizes the factorization scheme. In the steps we have outlined, we
have factored Figure \ref{fig:MPS-summary}(a) into the MPS in Figure
\ref{fig:MPS-summary}(d), in terms of the components listed below.  
\begin{enumerate}
 \item[(i)] States (labeled $\textbf{3}$, $\textbf{6}$ and
   $\textbf{9}$; denoted $\phi_3$, $\phi_6$ and $\phi_9$,
   respectively): $\phi_3=(\lambda_0,\lambda_1)$,
   $\phi_6=(\lambda_2,\lambda_3,\lambda_4,\lambda_5)$ and
   $\phi_9=(\lambda_6,\lambda_7)$.  The $\lambda_i$'s are the
   singular values across each partition.  The number of non-zero
   singular values ($\chi$) is given by the minimum dimension of the
   two halves from the cut.  For the case of qubits, the first outside
   partition has at most two non-zero entries, and the next inside
   partition has at most four.   One might also consider the singular
   values as the square roots of the eigenvalues of either member of the pair of reduced
   density matrices found from tracing out either half of a partition.
 \item[(ii)] Unitary gates (labeled $\textbf{2}$ and $\textbf{10}$; denoted $U_2$ and $U_{10}$, respectively).  
 \item[(iii)] Isometries (labeled $\textbf{5}$ and $\textbf{8}$;
   denoted $I_5$ and $I_8$, respectively).   The isometry condition
   describes the tensor relation $I_{jq}^d ~\overline{I}^{jq}_r =
   \delta ^d_{~r}$.  It is a consequence of the fact that tensors
   $I_5$ and $I_8$ arise from unitary gates, as explained in
   Remark~\ref{remark:isometry}.
   The isometry condition plays a more relevant role in structures
   other than 1D tensor chains.
\end{enumerate}

We note that by appropriately combining neighboring tensors as in
Figure~\ref{fig:MPS-summary}(a), one recovers 
the familiar matrix product representation of quantum
states~\ref{fig:MPS-summary}(b).  Matrix product states are written in
equational form as  
\begin{equation}\label{eqn:MPS}
 \ket{\psi} = \sum_{ijkm} A^{[1]}_i A^{[2]}_j A^{[3]}_k A^{[4]}_m\ket{ijkm}.
\end{equation}
Here $A^{[1]}$ becomes a new tensor formed from the contraction of tensors
labeled $\12$, $\13$, and $A^{[2]}$ is a 
contraction of tensors labeled $\15$ and $\16$, etc. 

\begin{remark}[Freedom in the representation]
Readers would have noticed that we made a choice to perform the
factorization starting from the left of the tensor and applying the SVD
successively on tensors as we moved to the right.  This apparent
ambiguity has been characterized in detail~\cite{2006quant.ph..8197P}. 
It corresponds to a gauge freedom (on internal wires) given by action of the
special linear group where the dimension of the representation is given by the
dimension of the wires (e.g.~the internal bond dimension).
 For open boundary conditions as has been considered here, there is a
``canonical gauge'' given first by Vidal.
It is unique up to degeneracies in
the spectrum of local reduced density operators~\cite{2006quant.ph..8197P}.
\end{remark}

A utility of our approach summarized in Figure \ref{fig:MPS-summary}(a) is that
the \COPY-tensor is well defined in terms of
purely graphical rewrite identities (as seen in Definition \ref{def:copy-prop}).
 These graphical relations allow one to gain insights (into e.g.\ polynomial
invariants as will be seen).
The factorization we present however, allows one to preform many diagrammatic manipulations with ease, and exposes more structure inherent in a MPS.  

\begin{remark}[Data compression]
 The compact representation of a MPS is recovered by picking a cutoff
 value for the singular values across each partition, or a maximum
 number of allowed singular values.  This allows one to compress data
 by truncating the Hilbert space and is at the heart of MPS computer
 algorithms in current use.
\end{remark}

The singular values found from the MPS factorization can be used to
form a complete polynomial basis to express invariant quantities related to an MPS.  

\section{Penrose notation meets entanglement invariants}
\label{sec:qubit-invariants}

Here we will consider the variant of the graphical tensor calculus of
Penrose~\cite{Penrose} we have tailored to represent and contract 
polynomial invariants.  
We must first recall the notions surrounding polynomial invariants.  

\subsection*{Polynomial invariants}
\label{sec:comdef}

Assume we are given a group~$G$, a vector space~$V$, and a
group representation $D: G \to \Aut(V)$.\footnote{
$\Aut(V)$ denotes the group of automorphisms of~$V$, \ie{} the
invertible linear maps from~$V$ to itself.}
Given a set~$Q$, a function $f: V \to Q$ is an \emph{invariant function} or simply
\emph{an invariant} under~$D$ iff it is
constant on the orbits of~$D$.\footnote{
Or equivalently iff $f$ itself is a fixed point under the induced
representation
$D': G \to \Aut(F(V, Q))$,
$D'(g)(f) = f \circ D(g^{-1})$.
}

In the context of quantum mechanics, the vector space~$V$ is typically
either the state space~$\hilb{H}$, or $\End(\hilb{H})$, the space of
linear operators~$\hilb{H} \to \hilb{H}$.
Any representation $D:~G~\to~\Aut(\hilb{H})$ on~$\hilb{H}$ induces a
representation $R:~G~\to~\Aut(\End(\hilb{H}))$ on~$\End(\hilb{H})$:
\be
R(g)(\rho) = D(g) \rho D^{-1}(g).
\ee


An important class of invariants are the \emph{polynomial
invariants} $f: V \to \C$, which are polynomial functions of the coefficients of 
$\rho$ or $\ket{\psi}$ in the standard basis.
The study of such polynomials is known as
invariant theory~\cite{Hilbert}.  David Hilbert made notable progress
on this topic, which he pursued throughout his
life.  There has been past work considering these invariants in the
context of quantum information science.
Some that was influential to us includes~\cite{grassl1998, makhlin2002, entinv3, qubits3, entinv1}.  See also the complementary recent study \cite{CM12}.  

\begin{remark}[Basis independence]
To form a polynomial out of the coefficients of a state, one first
chooses a basis to express the state in. The value of the
polynomial generally depends on the basis chosen.
However, a polynomial that is invariant under any group that contains the local
unitary group as a subgroup is inherently basis independent
as long as our basis is a tensor product of orthonormal local bases.
\end{remark}

\subsection*{Invariance under the local unitary group}

\begin{definition}[Local unitary (LU) equivalence of states]
Two quantum states (pure or mixed) in the Hilbert space
$\hilb{H} = \hilb{H}_1 \otimes \hilb{H}_2 \otimes \ldots \otimes \hilb{H}_n$
are LU equivalent iff they are related by a local
unitary transformation, that is, a member of the natural
representation of the group 
\be 
G_{\text{LU}} := U(1) \times SU(d_1)\times SU(d_2)\times \ldots  \times SU(d_n),
\ee
where $d_i = \dim \hilb{H}_i$~is the dimension of the $i$th subsystem.
LU equivalence yields a partitioning of the state space into LU orbits.
Entanglement measures are by definition LU
invariants, \ie, constant on the aforementioned equivalence classes.
\end{definition} 

We now present a diagrammatic method for systematically generating
polynomial LU invariants for state vectors and operators
by casting the method of Grassl et al.~\cite{grassl1998,Procesi76}
into a form based on the Penrose tensor calculus.
The method generates homogeneous polynomials in the state coefficients
that are necessarily invariants of the local
unitary group.\footnote{Although we can generate a complete set of invariants in this fashion,
except in rare cases, finding a \emph{minimal} complete set of polynomial
invariants is computationally difficult.  This alternative line of
research has been a key focus in the connection of invariant theory
with quantum entanglement.}  A utility of generating this set stems
from the fact that the tensor networks considered can be used to
calculate quantities that are invariant under the action of the local
unitary group.  

\begin{figure}[h!]
\includegraphics[width=0.95\textwidth]{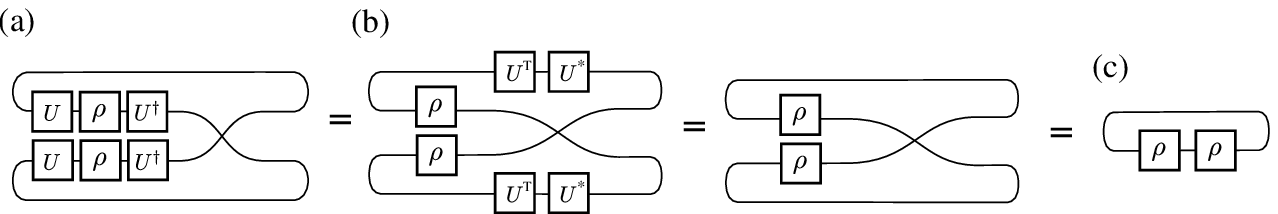}
\caption{Proof of the invariance of $I_{2; (12)}$.
Having acted on $\rho$ with some unitary operation $U$,
we slide $U$ and $U^\dagger$ around the bends, taking the transpose and resulting in (b).  The
unitaries cancel and the diagram reduces to (c), showing that it
indeed describes an invariant. A little bit of further
manipulation shows that $I_{2; (12)}$
evaluates to $\Tr(\rho^2)$.
\label{fig:invproof}
}
\end{figure}

Given a density operator $\rho: \hilb{H} \to \hilb{H}$, consider the network 
\be
 \includegraphics[width=0.3\diagwidth]{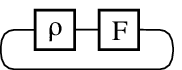}
\ee
equivalent to the expression
$\Tr(F \rho) = F^i_{~j} \rho^j_{~i}$. By choosing a suitable~$F$, we
can represent any first-degree homogeneous polynomial in the
coefficients of~$\rho$ in this way.
Likewise, the tensor network
\be
 \includegraphics[width=0.3\diagwidth]{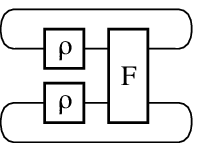}
\ee
translates to
$\Tr(F \rho^{\otimes 2}) = F\indices{^{il}_{jk}} \rho\indices{^j_i} \rho\indices{^k_l}$,
giving all the second-degree homogeneous polynomials.
The procedure carries on in this fashion. A diagram with~$k$
copies of~$\rho$ gives us all the homogeneous polynomials of degree~$k$:
\be \label{eqn:F}
\Tr(F \rho^{\otimes k}) = F\indices{^{ij \cdots m}_{pq \cdots t}}
\rho\indices{^p_i} \rho\indices{^q_j} \cdots \rho\indices{^t_m}.
\ee 
Having thus generated a complete basis for the polynomials in the
coefficients of~$\rho$, we next wish to find out which of these
homogeneous polynomials are invariant under the natural representation of~$G_{\text{LU}}$:
\begin{equation}
\label{eqn:density-invariant} 
\Tr(F (U \rho U^{-1})^{\otimes k})
= \Tr((U^{-1})^{\otimes k} F U^{\otimes k} \rho^{\otimes k})
= \Tr(F \rho^{\otimes k}) \qquad \forall U \in G_{\text{LU}}, \quad
\forall \rho.
\end{equation}  
This is fulfilled iff
\be 
[F, U^{\otimes k}] = 0 \qquad \forall U\in G_{\text{LU}}.
\ee 
We are then faced with finding matrices $F$ that commute with
$U^{\otimes k}$ for each $U\in G_{\text{LU}}$. The solution is roughly stated in
the following theorem.
\begin{theorem}[Brauer~\cite{brauer}, Procesi \cite{Procesi76}]\label{theorem:brauer}
The algebra of matrices that commute with every $U^{\otimes k}$ for $U \in G_{\text{LU}}$ is
generated by the unitary representation
\be
T: \underbrace{S_k \times \ldots \times S_k}_{n \: \text{copies}} \to \Aut \hilb{H}^{\otimes k}
\ee
of the $n$-fold direct product of the
permutation group~$S_k$ which, independently for each of the~$n$ subsystems,
permutes the relative ordering of the $k$~copies of that subsystem's
state space within the total space~$\hilb{H}^{\otimes k}$.
\end{theorem}
Hence, it is enough to consider matrices~$F$ which correspond to these
permutation maps.
The permutation group has a well known and evident diagrammatic
form. Below, we show the elements of the permutation group (a) $S_1$,
(b) $S_2$, and (c) $S_3$.
\be
 \includegraphics[width=1.5\diagwidth]{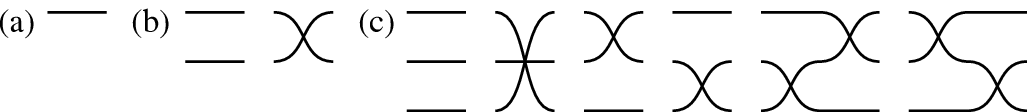}
\ee
We then carry on to evaluate all the expressions of the form
\be
\label{eqn:I} 
I_{k; \: \sigma_1, \sigma_2, \ldots, \sigma_n}(\rho) := \Tr(T(\sigma_1, \sigma_2, \ldots, \sigma_n) \rho^{\otimes k}), \quad
\text{where} \: \sigma_i \in S_k
\ee
to generate the homogeneous invariant polynomials of
degree~$k$.\footnote{
We use the cycle notation to denote specific elements~$\sigma$ of the
permutation groups.
}
It is easy to see that the diagrams we obtain are indeed invariant
under~$G_{\text{LU}}$, as shown in \Figref{fig:invproof}.

Not all of these invariants are independent, or even distinct.
We can eliminate some of the redundancy using the following theorem, proven in~\cite{grassl1998}:
\begin{theorem}[Invariant distinctness~\cite{grassl1998}]
\label{th:distinctness}
Since all the copies of~$\rho$ in \eqref{eqn:I} are identical, we
may permute their relative order without changing the invariant.
This is equivalent to conjugating each subsystem
permutation~$\sigma_i$ with the same element~$\tau \in S_k$:
\begin{align}
I_{k; \: \sigma_1, \ldots, \sigma_n}
= I_{k; \: \tau \sigma_1 \tau^{-1}, \ldots, \tau \sigma_n \tau^{-1}} \qquad \forall \sigma_i, \tau \in S_k.
\end{align}
\end{theorem}
This theorem enables us to arrange each invariant diagram to the
following canonical form which makes it easy to tell if two diagrams
are topologically distinct.
\begin{enumerate}
\item[(i)]
The $k$ copies of the system are arranged such that
the permutation on the first subsystem is grouped by cycles, ordered
by non-increasing cycle length.

\item[(ii)] 
The process is repeated on the second, then third etc. subsystem within the
remaining permutational freedom, i.e.~cyclic permutation within the
cycles and permuting cycles of identical length.
\end{enumerate}
If a particular diagram is not connected, the corresponding invariant
is the product of the invariants corresponding to the
disjoint sub-diagrams.
\begin{remark}
Note that using the procedure here, two algebraically independent invariants necessarily have topologically
distinct diagrams but the converse does not necessarily hold.
\end{remark}

\begin{theorem}[Real-valuedness of the invariants]
If all the permutations~$\sigma_i$ are self-inverse, or can all be
inverted by conjugating them with the same element~$\tau$ as shown in
Theorem~\ref{th:distinctness}, the invariant
$I_{k; \: \sigma_1, \ldots, \sigma_n}$ is necessarily real for all states.
\begin{proof}
\begin{align}
\notag
I^*_{k; \: \sigma_1, \ldots, \sigma_n}(\rho)
&= \Tr((\rho^{\otimes k})^\dagger T^\dagger(\sigma_1, \ldots, \sigma_n))
= \Tr(T^\dagger(\sigma_1, \ldots, \sigma_n) \rho^{\otimes k})\\
&= \Tr(T(\sigma_1^{-1}, \ldots, \sigma_n^{-1}) \rho^{\otimes k})
= I_{k; \: \sigma_1^{-1}, \ldots, \sigma_n^{-1}}(\rho).
\end{align}
\end{proof}
\end{theorem}

\begin{theorem}[States with a single subsystem]
All the independent invariants of a $d$-dimensional state~$\rho$ with a single
subsystem are given by
\be
I_k := \Tr(\rho^k), \quad \text{where} \quad k \in \{1, 2, \ldots, d\}.
\ee
\begin{proof}
The only degree one invariant, $I_1~:=~\Tr(\rho)$, is presented in
\Figref{fig:inv_n1}.a.
The two possible degree two invariants are shown in
\Figref{fig:inv_n1}.b. The first one is simply~$I_1^2$.
The second one, $I_2~:=~\Tr(\rho^2)$, however is independent.
Likewise, the only independent invariant of degree three,
$I_3~:=~\Tr(\rho^3)$, is given in \Figref{fig:inv_n1}.c.
In general, at each degree~$k$ we obtain a single new independent
invariant~$I_k~:=~\Tr(\rho^k)$,
by using a complete permutation connecting all the~$k$ copies of the state.
\begin{figure}[h!]
\includegraphics[width=1.0\diagwidth]{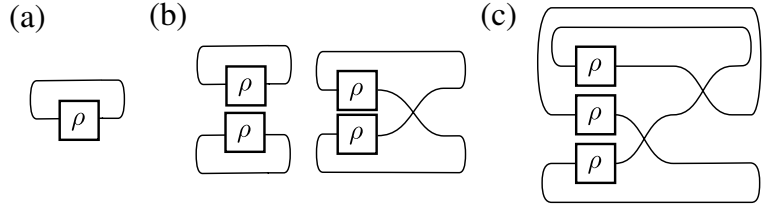}
\caption{LU invariants for a single subsystem.
(a) $I_1 := I_{1; e} = \Tr(\rho)$.
(b) $I_{2; e} = I_1^2$. $I_2 := I_{2; (12)} =\Tr(\rho^2)$.
(c) $I_3 := I_{3; (123)} = \Tr(\rho^3)$.
\label{fig:inv_n1}
}
\end{figure}
 
The Cayley-Hamilton theorem now tells us that the basis is finitely generated, as
every~$\rho$ satisfies its own characteristic polynomial, giving an $d$th degree
polynomial equation in~$\rho$ which enables us to express any~$I_m$
with~$m > d$ in terms of the lower-degree invariants~\cite{Procesi76}.
\end{proof}
\end{theorem}

\begin{example}[Invariants for a single qubit]
The only independent (fundamental) invariants of a single qubit state are~$I_1$
and~$I_2$, defined in the previous theorem.
$I_1 = \Tr(\rho)$ is the norm of the state.
$I_2 = \Tr(\rho^2)$
turns out to be precisely the purity of the state.
In terms of the eigenvalues $(\lambda_0, \lambda_1)$ of~$\rho$ we have 
$I_1 = \lambda_0 + \lambda_1 = 1$ (for normalized states)
and $I_2 = \lambda_0^2 + \lambda_1^2$.
From the Cayley-Hamilton theorem we have that there is a second
degree monic polynomial in~$\rho$ that vanishes identically.
In other words, constants $a, b$ exist such that 
\be 
\rho^2 + a \rho + b \I = 0.
\ee
Multiplying both sides by~$\rho^m$ and taking the trace, we obtain the
recurrence relation
\be
I_{m+2} +a I_{m+1} +b I_{m} = 0,
\ee
and thus find that the traces of
higher powers of $\rho$ can be expressed in terms of $I_1$ and
$I_2$.
These invariants are indeed algebraically independent and complete,
meaning any other polynomial invariant can be expressed in
$\{\R, +, \cdot, I_1, I_2\}$.  For instance, 
\be
\det(\rho)
= \frac{1}{2}\left(\Tr(\rho)^2 - \Tr(\rho^2) \right)
= \frac{1}{2}\left(I_1^2 - I_2\right)
= \lambda_0 \lambda_1.
\ee 
Likewise, $I_3 = \lambda_0^3 + \lambda_1^3$ can be written as 
\be 
I_3 = I_1(I_2 - \det(\rho))
\ee 
\end{example} 

\begin{remark}[Bipartite states]
For bipartite states we obtain a much more complicated set of
invariants.
\Figref{fig:inv_n2} presents all the topologically distinct invariants up to~$k=3$.  
\begin{figure}[h!]
\includegraphics[width=0.9\textwidth]{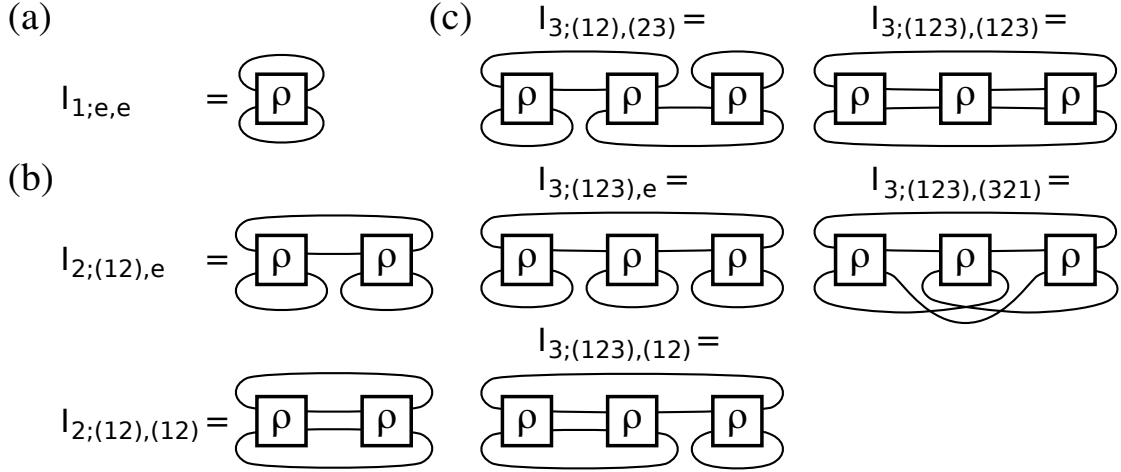}
\caption{LU invariants of a bipartite system up to~$k=3$.
To avoid listing essentially similar diagrams we only show here
the distinct invariants modulo swapping the order of the two subsystems.
(a)~The only first degree invariant is $I_{1; e, e} = \Tr(\rho)$, same as
with a single subsystem.
(b)~In the second degree we obtain a new invariant,~$I_{2; (12),e}$.
(c)~There are several new third degree invariants, including the
topologically distinct $I_{3; (123), (123)}$ and $I_{3; (123), (321)}$.
\label{fig:inv_n2}
}
\end{figure}
\end{remark}

\subsection*{Pure states}

If the state~$\rho$ is pure, the diagrammatic structure of the LU invariants
simplifies considerably, and many of the diagrams break up into
unconnected sub-diagrams. Furthermore, in the case of bipartite pure states, we may
apply the Schmidt decomposition and introduce graphical rewrite rules to show that these invariants reduce to
polynomials of the Schmidt coefficients.

\begin{figure}[h!]
\includegraphics[width=0.7\textwidth]{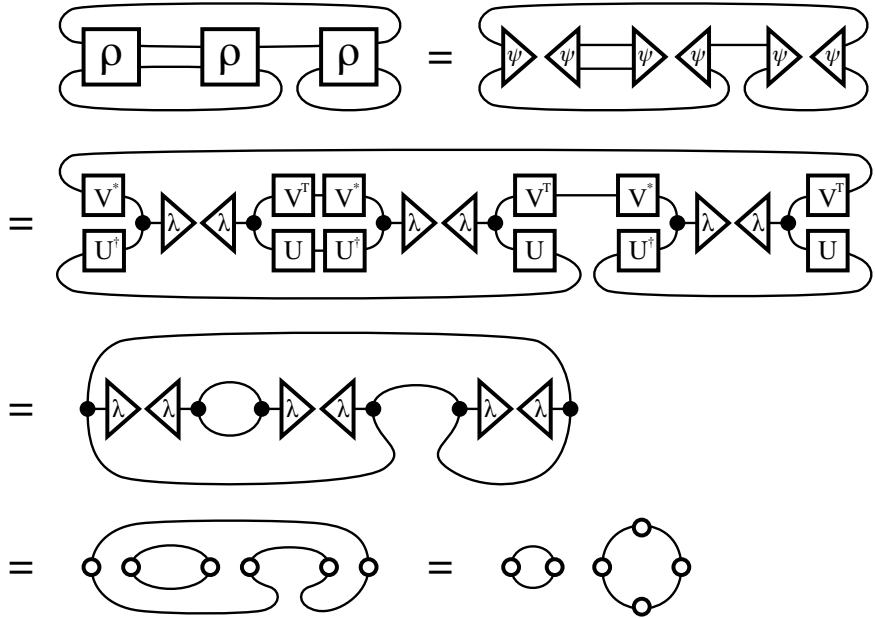}
\caption{Invariant $I_{3; (123),(12)}$ evaluated for a pure
bipartite state $\rho = \ketbra{\psi}{\psi}$ using the diagrammatic Schmidt decomposition.
The unitaries~$U$ and~$V$ (and possible dimension changers~$Q$) cancel, and one is left with two disjoint loops, on which
the blank circles denote diagonal tensors with the Schmidt
coefficients~$\{\sigma_i\}_{i=0}^{d-1}$ on the diagonal.
\label{fig:invariant_schmidt}
}
\end{figure}

\begin{theorem}[Bipartite pure states]

Applying the diagrammatic Schmidt decomposition
presented in Corollary~\ref{ex:diagrammatic-Schmidt} to
the bipartite invariant
diagrams in~\Figref{fig:inv_n2},
we can see that the unitaries~$U$, $V$ and the dimension changers~$Q$
always cancel, and the invariant diagrams break up into mutually
disjoint loops corresponding to
sums of even powers of the Schmidt coefficients~$\{\sigma_i\}_{i=0}^{d-1}$.
Hence, the only independent invariants we obtain are of the form
\be
J_{k} := I_{k; \; (12 \cdots k), e} = \sum_i \sigma_i^{2k} \quad
\text{for} \quad k \in \{1, 2, \ldots\}. 
\ee
\end{theorem}
In \Figref{fig:invariant_schmidt} we present this process for the invariant
\be
I_{3; (123),(12)}
= \left(\sum_k \sigma_k^2\right) \left(\sum_k \sigma_k^4\right)
= I_{1; \; e,e} \: I_{2; \; (12),e}
= J_1 \: J_2.
\ee

Since the $d$ Schmidt coefficients themselves (by construction) form a complete set of
bipartite LU invariants, we should be able to express them as
functions of~$\{J_i\}_{i=1}^{d}$. This is accomplished in principle
by solving the following system of polynomial equations:
\begin{align}
\notag
\sum_i \sigma_i^2   &= J_1,\\
\notag
\sum_i \sigma_i^4 &= J_2,\\
\notag
\vdots \quad &= \: \vdots\\
\sum_i \sigma_i^{2d} &= J_d.
\end{align}

\begin{example}[Pure state of two qubits]\label{ex:qubits} 
Given a pure two-qubit state
\be 
\ket{\psi} = \sum_{ij} \alpha^{ij} \ket{ij},
\ee
we wish to compute its LU invariants.
Since it has two Schmidt coefficients, we expect to
find two independent invariants.
The first one,~$J_1$,
corresponds to the squared norm of the state, and is thus trivially
invariant under unitary transformations of~$\ket{\psi}$:
\be 
\includegraphics[width=.8\textwidth]{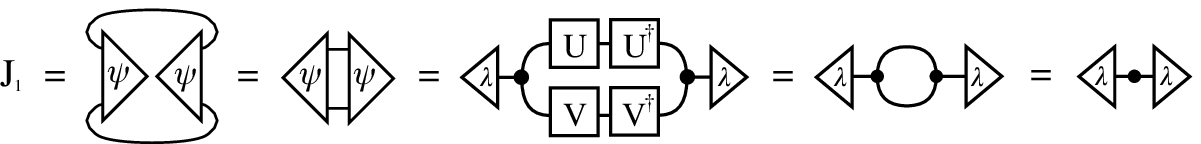}
\ee
The second independent invariant,~$J_2$, doesn't have as simple an
interpretation:
\be
\includegraphics[width=.85\textwidth]{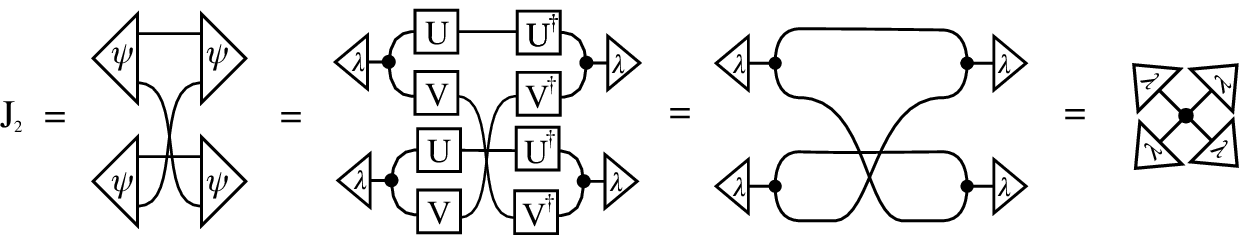}
\ee 
The diagrams correspond to the equations
\begin{align}
J_1 &= \sum_{ij} \alpha^{ij} \alpha_{ij}^* = \sigma_0^2 + \sigma_1^2,\\
J_2 &= \sum_{ijkl} \alpha^{i j}\alpha^{k l} \alpha_{i l}^* \alpha_{kj}^* 
= \sigma_0^4 + \sigma_1^4,
\end{align}
with the solution
\be
\sigma_{0,1}^2 = \frac{1}{2}\left(J_1 \pm \sqrt{2 J_2 -J_1^2} \right),
\ee
which yields $J_2 \le J_1^2 \le 2 J_2$. 
For a normalized state ($J_1=1$),
it can be shown that the invariant $J_2$ can be expressed as 
\begin{equation} 
J_2 = 1-2 |\alpha_{00} \alpha_{11} -\alpha_{01} \alpha_{10}|^2, 
\end{equation} 
where $\alpha_{00} \alpha_{11} -\alpha_{01} \alpha_{10}$ is simply 
the determinant of the coefficient matrix~$\alpha$,
which is non-zero iff $\ket{\psi}$ represents an entangled state.

$J_1$ and $J_2$ are the only algebraically independent LU invariants
of a pure two-qubit system.
Any polynomial function of such invariants is also a polynomial
invariant.  In this fashion, it is a remarkable feature that functions
of $J_1$ and $J_2$ are all that is needed to express any local unitary
invariant of two-qubit pure states.  This elementary result follows from a
much more powerful and general result in classical invariant theory,
a proof by Hilbert that the ring of polynomial invariants is
finitely generated~\cite{Hilbert}.  This corresponds to freely
generated linear sums and products of $J_1$, $J_2$, e.g. the ring $\{J_1,
J_2, (\R, +, \cdot)\}$.  Any minimal complete set of invariants that
can freely generate the full ring are called \emph{fundamental invariants}.
\end{example}
%
%

\section*{Invariants, entropies and entanglement}
\label{sec:entanglement} 

Here we focus on expressing Rényi entropies in terms of the 
invariants we have found tensor contractions for in the previous sections.  The Rényi entropy has many uses in condensed matter physics (see for example \cite{2009PhRvL.103z1601F, 2010PhRvL.104o7201H}) and recently has been given an interesting physical interpretation \cite{Renyi}.  We will recall the definition as 
\begin{definition}[Rényi entropy \cite{Renyi2}]\label{def:reyni}
 The Rényi entropy of order $\alpha$ is defined as 
 \be 
S_\alpha := \frac{1}{1-\alpha} \ln \Tr (\rho^\alpha).
 \ee  
In the limit $\alpha \to 1$ we obtain
\be 
\lim_{\alpha \to 1}S_\alpha = -\Tr(\rho \ln \rho),
\ee
which is the von Neumann entropy.
\end{definition}

Here we note that terms such as $\Tr (\rho^\alpha)$ are in
correspondence with the tensor contractions evaluating to invariants
which we have already found.  To explain how we can contract tensor
networks to evaluate Rényi entropies for counting $\alpha>1$, we will
close the paper with a specific example, though the procedure we
describe is general.  We will focus explicitly on the invariants of a
bipartition of a 5-party qubit state $\ket{\psi}\in \7 C^2\otimes \7
C^2\otimes \7 C^2\otimes \7 C^2\otimes \7 C^2$.  We first recall that
we can factor any state into a MPS; in the case of our example, this
yields the following graphical depiction.
\be
 \includegraphics[width=.9\textwidth]{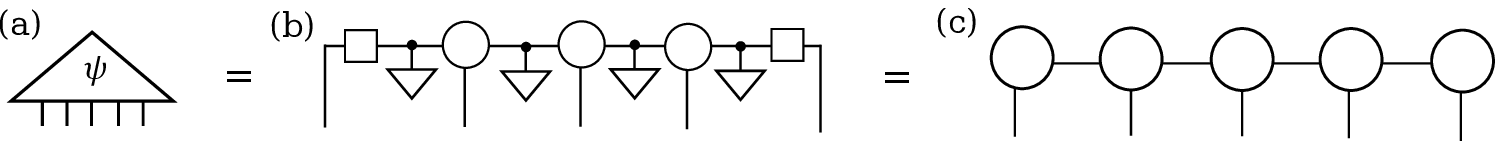}
\ee
Here (a) is the original state, (b) is our factorization in terms of \COPY-tensors, and (c) recovers the familiar MPS representation, as explained in Section \ref{sec:dmps}.  The method to evaluate Rényi entropies by tensor contraction works generally by first grouping the legs of any tensor network state into a bipartition, and to consider the correlations between the two halves.  In the present example, we group the two top legs ($A$) and the other three legs ($B$) and then apply the graphical SVD:  	
\be
 \includegraphics[width=.75\textwidth]{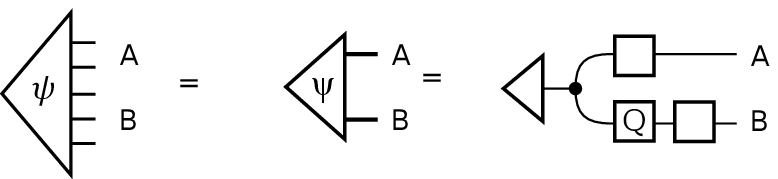}
\ee
After tracing out system $B$, we are concerned with a four dimensional space $\7 C^2\otimes \7 C^2$.  From the Cayley-Hamilton theorem 
\begin{eqnarray}\label{eqn:cayley}
\Tr(\rho ^4) + a \Tr(\rho^3) + b \Tr(\rho^2) + c &=& 0
\end{eqnarray} 
for some values of $a,b,c$.  We hence conclude that all information we can expect to find can be obtained by evaluating tensor contractions for $\Tr (\rho^n)$ for $n=2,3,4$ as given in the section on invariants.  As an example, to evaluate $\Tr(\rho^3)$: 
\be
 \includegraphics[width=.5\textwidth]{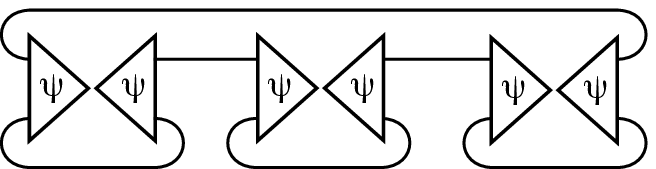}
\ee
We can evaluate this contraction using a tensor network numerical algorithms package.  By writing it in terms of the graphical SVD and applying the rewrite rules developed in the present work, we arrive at 
\be
 \includegraphics[width=\textwidth]{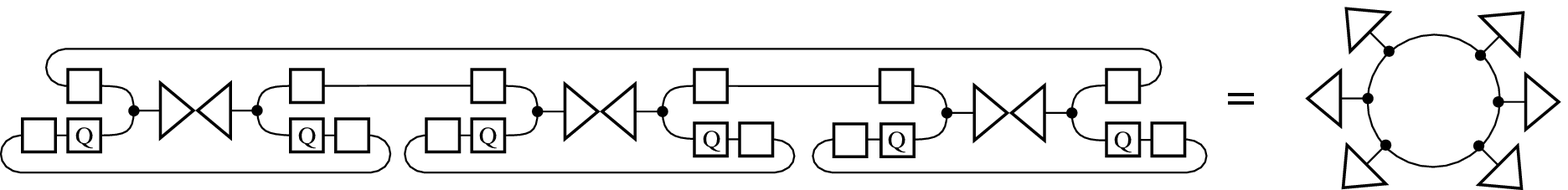}
\ee
This illustrates that the network reduces to an expression in terms of
the singular values of the pure state.
If the singular values of the original state are given as
$(\sigma_1, \sigma_2, \sigma_3, \sigma_4)=:(\sqrt p_1, \sqrt p_2, \sqrt p_3, \sqrt p_4)$,
then the expression evaluates to 
\begin{equation} 
I_{3; (123), e} =  \sigma_1^6 + \sigma_2^6 + \sigma_3^6 + \sigma_4^6 = p_1^3 + p_2^3 + p_3^3 + p_4^3 
\end{equation}
where the eigenvalues $p_i$ on the right are the probabilities of measuring the
reduced system in the $i$th eigenstate.  We then express the Rényi entropy with
$\alpha = 3$ as $S_3=-\frac{1}{2} \text{ln}(I_{3;(123),e})$. Other quantities can be
similarly calculated, resulting in the following identical relation for the Rényi entropies
\begin{equation} 
\exp(-3 S_4) + a \exp(-2 S_3) + b \exp(-S_2) + c = 0 
\end{equation} 
which is an alternative expression for \eqref{eqn:cayley} in terms of the $S_\alpha$ from Definition \ref{def:reyni}. 

\section{Conclusion}
We have developed a graphical method for expressing a complete polynomial
basis for the local unitary invariants of any finite-dimensional quantum system.
Using the diagrammatic SVD, we
have shown that for pure bipartite systems
these contractions can be expressed in terms of manifestly
invariant singular values.  These methods seem to provide new
conceptual insight to understanding quantities such as entropies which
are expressed as holomorphic functions of the invariants.  By
connecting invariant theory with tensor network states, we hope that
this work leads to a better understanding of how entropies and entanglement 
measures can be calculated for specific models, and ultimately to a better
understanding of invariants in various higher-dimensional geometries
as well.

\begin{acknowledgments}
We thank Markus Grassl, Ann Kallin, Jason Morton, Nilhan Gurkan and Francesco Vaccarino.  VB visited the Centre for Quantum Technologies (CQT, Singapore),  
ML visited Oxford University and JB visited the Perimeter Institute for
Theoretical Physics as well as the Institute for Quantum Computing during part of this work. 
\end{acknowledgments}

\bibliography{qc}

\end{document}